# Understanding working time and relocation choices of ridehailing drivers[1]

Yuanjie Tu[a*], Moein Khaloei[a], Natalia Zuniga-Garcia[b], Don MacKenzie[a]

[a] Department of Civil and Environmental Engineering, University of Washington, 119 More Hall, Seattle 98195, US

[b] Argonne National Laboratory, 9700 South Cass Avenue, Lemont, Illinois 60439, US

*Corresponding author Email: yuanjt2@uw.edu

**ABSTRACT:** We identified four types of ridehailing drivers and jointly modeled driver working time and relocation choices using a stated preference survey of 200 drivers in Seattle, US. In the working time choice level, our finding mainly echoes the neo-classic theory of driver labor supply: all types of drivers choose to continue working as the hourly earning rate increases. We see little evidence of an earnings threshold effect as the reference dependent theory states, even when the driver reported an earnings target in the survey. However, there is evidence of reference dependent behavior with respect to the working time variable: *time-target drivers* and *both-target drivers* are less likely to continue working when they hit their working time targets. Relocation choice is mainly impacted by surge price, average trip waiting time, and relocation time: higher surge price in the current neighborhood encourages the driver to stay, while higher surge price in a nearby neighborhood attracts the driver to relocate. Drivers are more likely to stay in a neighborhood where the average trip waiting time is low. Longer relocation time discourages drivers to relocate. In addition, drivers are more likely to stay in the same place, everything else being equal.

---

[1] Paper major revision at *Transportation*

# 1. Introduction

As ridehailing services have surged in popularity, the behaviors of drivers have received much less attention than those of passengers. However, few have recognized that a sufficient driver supply[2] is the foundation of high-quality ridehailing services and have focused their research accordingly (Ashkrof et al. 2020). From the supply side, drivers are not only chauffeurs but independent contractors, which means transportation network companies (TNCs) that match drivers with passengers are limited in their ability to dictate drivers' actions (Wentrup et al. 2018). Drivers provide their vehicles to transport passengers and make their own work decisions such as when to stop driving, where to relocate, and whether to accept the trip request. Insufficient driver supply leads to lower ridehailing service quality (longer waiting times and higher prices per ride), which damages passenger satisfaction and could further undermine TNCs' ability to fulfill urban mobility needs.

The COVID-19 pandemic has furthermore put this question front and center. The cost of a ride increased 92% between January 2018 and July 2021, along with longer waiting times since many drivers decreased their daily working time, avoided certain areas in the city, and/or left the ridehailing market (Evans 2021). Recently, TNCs have been finding it difficult to lure drivers back, work longer hours, or serve particular areas, which leaves increasingly more ridehailing demand unmet (Bursztynsky 2021). In this sense, a sufficient driver supply is even more important during crisis than normal times, in helping cities maintain resiliency – the capability to adapt to and recover from stresses, shocks, and hazards (National Research Council 2012; Valdés Cano 2021).

This study jointly models two ridehailing driver behaviors that are central to driver supply: working time and relocation choices. These two behaviors impact driver supply through time and space: drivers can freely decide when and where to provide their labor. Relying on data from a stated choice experiment, we jointly estimate models of drivers' choices about whether to continue working and whether to relocate while working. While driver working time choices affect the overall driver supply and service quality at a given time, their relocation choices affect the distribution of service quality across different areas of a city. We modeled these two behaviors jointly because they are so intertwined that it is almost impossible to think about one without the other: a driver can only choose where to go if they first choose to continue working; a driver's decision whether to continue working depends in part on the expected rewards in working, which in turn depends on the level of demand in the current location and nearby locations. The model we propose can serve as a tool for transport agencies to predict and adjust future ridehailing driver supply.

---

[2] Ridehailing driver supply means the number of drivers available to provide ridehailing services.

## 2. Literature review

In this section, we present a summary of existing studies on drivers' working time and relocation choice. We also summarize the current research gaps.

### 2.1 Working time choice

Drivers' working time choice is not a new question. There has been a vast literature in economics discussing drivers' working time choice starting from the 20$^{th}$ century, represented by the term "driver labor supply" (Camerer et al. 1997).

Among the investigators who have examined drivers' working time, there is a heated argument: the neo-classical theory of labor supply states that the quantity of labor supplied increases as the wage increases, with drivers working longer hours on days when they are earning more per hour (Lucas and Rapping 1969). The reference dependent utility theory of labor supply suggests that the relationship is negative, as drivers tend to stop working once they reach a reference or target level of daily earnings (Tversky and Kahneman 1991). A large pool of literature has tried to empirically test the two hypotheses, but the results are mixed. Camerer et al. (Camerer et al. 1997) and Farber (Farber 2005) used the same trip sheet dataset of New York taxi drivers and came to different conclusions. Camerer et al. (Camerer et al. 1997) found that higher earning rate, defined by daily earnings divided by daily working time, is associated with shorter expected daily working time. This result means taxi drivers might have a daily earnings target and would stop driving after the target is reached. Farber (Farber 2005), on the contrary, supports the neoclassical theory. He pointed out several econometric weaknesses of Camerer et al. (Camerer et al. 1997) and concluded that taxi drivers' probability of stopping driving mainly depends on accumulated working time that day, but not necessarily accumulated earnings. He concludes that for taxi drivers, "tomorrow is another day", so they do not worry about setting a target for a single day. The difference between Camerer et al. (Camerer et al. 1997) and Farber (Farber 2005) are mainly in their conceptual frameworks and model specifications. The intensified debate can be seen across the massive literature on this topic (Agarwal et al. 2013; Crawford and Meng 2011; Stafford 2015; Xu et al. 2020). – extend a little bit.

Three main limitations exist in current research on drivers' working time choice. First, most work has focused on taxi drivers (Agarwal et al. 2013; Camerer et al. 1997; Crawford and Meng 2011; Farber 2008; Farber 2005), while few studies have paid attention to ridehailing drivers (Angrist et al. 2021; Chen et al. 2019; Sun et al. 2019; Xu et al. 2020). Nevertheless, ridehailing and taxi drivers differ in some important ways: First, ridehailing drivers are more flexible in their working time choices than taxi drivers because they are not constrained by "switching shifts" (Ashkrof et al. 2020; Chen et al. 2019). Second, the TNC platform matches ridehailing drivers with passengers more efficiently than the traditional taxi industry, which may lead to less waiting time and a higher earning rate (Brown and LaValle 2020; Schwartz 2018). Third, the compensation models for ridehailing and taxi drivers are different. Taxi drivers pay a fixed payment independent of how much they earn to the company, while ridehailing drivers pay a proportion of their trip fares (Angrist et al. 2021). Considering these differences, ridehailing drivers' working time choices might differ significantly from those of taxi drivers.

A second limitation of existing studies is that they have investigated the effects of earnings targets using observational data, without explicitly measuring drivers' earnings targets (Farber 2008; Xu et al. 2020). While there are some advantages using observational data, doing so increases the difficulty of investigating the effects of targets since those targets are not directly observed.

Finally, building on the above point, many current studies have discussed the effects of earnings targets, but none have considered time targets. Most research assumes that drivers will be less willing to continue working as their cumulative time worked increases, but no studies have discussed the possibility that drivers also have a reference level for working time.

## 2.2 Relocation choice

Many studies have investigated the spatial distribution of ridehailing trips (Barajas and Brown 2021; Marquet 2020; Shokoohyar et al. 2020; Wang and Mu 2018). However, similar though they might they seem, ridehailing drivers' relocation choices are different than the spatial distribution of ridehailing trips. The spatial distribution of ridehailing trips is the result shaped by both ridehailing supply and demand, while ridehailing drivers' relocation choices only represents the supply in the space dimension (though relocation choices are, as we show, affected by the levels of demand in various locations). Therefore, understanding the spatial distribution of ridehailing trips does not by itself provide a complete understanding of ridehailing driver supply (Hassanpour et al. 2020).

There are two main strands of studies on ridehailing drivers' relocation choices. One strand of studies is dedicated in optimizing the efficiency of drivers' searches for passengers so as to promote service quality (Afeche et al. 2018; Danassis et al. 2020) to reduce traffic congestion (Kim et al. 2020). A common limitation of this strand of studies is that they tend to assume driver behaviors are fully compliant with the platform and neglect the fact that drivers are independent contractors who freely decide where to go.

Another strand of studies focuses on strategically maximizing driver earnings (Henao and Marshall 2019; Zuniga-Garcia et al. 2020). Current studies suggest that in general, drivers earn more when they park and wait for the next trip request rather than drive to a more active location. The situation is only different if doing so saves at least 30% of trip waiting time (Henao and Marshall 2019). While strategically maximizing the earnings seem "ideal", drivers in the real life might have their own decision-making mechanisms.

Few peer-reviewed studies have modeled ridehailing drivers' relocation behaviors (Ashkrof et al. 2020). Ashkrof et al. (Ashkrof et al. 2020) qualitatively investigated ridehailing drivers' relocation strategies. They found that new drivers would prefer to move around, while experienced drivers would prefer waiting in the same location to get a ride. They also found that new drivers said they would choose to chase the surge, while more experienced drivers said they would not. To our knowledge, there are no studies quantitatively modeling ridehailing drivers' relocation behaviors and related factors.

## 2.3 Contribution

The key observations from this vast literature are four-fold. First, current studies model drivers' working time and relocation choices separately. However, drivers' working time and relocation choices are inextricably bound up together. A driver can only choose where to go if they choose to continue working; and a driver decides whether to continue working in part based on the expected rewards in working, which in turn depend on the levels of demand in the current location and nearby locations. Second, current studies on working time choice mainly focus on taxi drivers, with little attention on ridehailing drivers. Third, existing studies have not measured drivers' targets explicitly. Fourth, there is little work on descriptively and quantitatively modeling ridehailing drivers' relocation choices and related factors.

The present study is the first to our knowledge that models ridehailing drivers' working time and relocation choices jointly. We rely on a survey of Seattle area ridehailing drivers that included a stated choice experiment and explicitly asked drivers about their working time and earnings targets. This allows us to understand different working time decision-making mechanisms of different type of drivers. Our study adds to the literature by filling the aforementioned research gaps, and the result of our study can serve as a tool for transport agencies to predict and adjust future ridehailing driver supply.

## 3. Data collection

Our primary data source is a survey of 200 ridehailing drivers in the Seattle, USA region. Two advantages stand out when using a stated preference survey in this context. First, it allows us to directly capture drivers' daily working time and earnings targets, which will furthermore help us to investigate the relationship between drivers' working time choices and these targets. Second, it enables us to investigate the causal relationship between working time choice, relocation choice, and driving characteristics we are interested in. The inclusion of a choice experiment in the survey allows us to control endogeneity, a common problem that can lead to omitted variable bias when using non-experimental data (Abdallah et al. 2015).

The survey was conducted from August 11 to September 9, 2021. We trialed online and in-person data collection approaches. We finally adopted an in-person computer assisted personal interview approach because we found that (1) the response rate to an online version of the survey was very low, and (2) many online respondents appeared not to be real ridehailing drivers, based on screening and quality check questions in the survey. More details on the performance of different data collection approaches can be found in (Tu et al. 2021).

We recruited participants and assisted them to complete the survey on an iPad in the Seattle-Tacoma International Airport ridehailing driver waiting area. Respondents were compensated with an $15 Amazon gift card in the indicated amount.

Anecdotally, ridehailing drivers do not seem to trust surveys and interviews. The most frequently asked questions when administering the survey were "what is the purpose of the study?" and "do you work for Lyft or Uber?". Their major concern was that the responses would be used against them by ridehailing companies. This echoes existing evidence that drivers and companies have a tense relationship (Ashkrof et al. 2020). A total of 200 responses were collected from approximately 250 drivers who were invited to participate.

### 3.1 Online survey design

We conducted a series of pilot interviews with ridehailing drivers before designing the questionnaire, to ensure we captured the most important variables and used terminology familiar to the drivers. The questionnaire contained three sections: basic driving information, choice experiments, and background information.

*Basic driving information*

We asked respondents to provide information on which ridehailing companies they are driving for, how long they have been a ridehailing driver, whether they have another job besides ridehailing driver, the time they generally start and stop driving each day, any targets for working hours or earnings per day, and the numbers of trip requests they (1) received and (2) rejected in the past week.

To capture respondents' working time targets, we first ask them "Do you usually try to work a certain number of hours per day?" If their answer is "yes", we asked them "How many hours do you usually try to work per day?" Similarly, we captured their earnings target by first asking them "Do you usually try to earn a certain amount of money per day?" If they answered "yes", we asked them "How much do you usually try to earn per day?"

After capturing working time targets and earnings targets, we categorized driver respondents into four types based on whether they reported having targets of one or both types: *no-target drivers*, *time-target drivers*, *earnings-target drivers*, and *both-target drivers*. These driver types will be further used in our modeling work.

*Choice experiments*

To investigate the causal relationship between the drivers' working time choices, relocation choices, and driving characteristics, we adopted a blocked factorial design. The experimental design was conducted as follows. First, we chose 11 variables as experimental variables, including daily earnings, working time, ridehailing demand (represented as a combination of surge price and expected waiting time for a ride request) in the current neighborhood and four nearby neighborhoods, and the relocation time to drive to the four nearby neighborhoods. We include these variables because they are the most important variables reported by drivers in the pilot interviews. Experimental variable descriptions are shown in Table 1. Second, we generated levels for each experimental variable (Table 2). Third, we generated the full matrix of combinations of the variable levels. Next, we created blocks of six combinations (choice scenarios) for each respondent. In each scenario, she/he was asked to first choose to "stop working" or "continue working" based on different combinations of experimental variable levels; if they chose to continue working, they were asked to choose a neighborhood from among the current neighborhood they are in and four nearby neighborhoods. The four nearby neighborhoods were unlabeled alternatives, i.e. they were not identified as specific neighborhoods in the Seattle region. Figures 1 and 2 show an example of driver working time and relocation experiment.

*Background information*

In this section, we asked respondents questions about their socio-demographics including age, gender, race, whether born in the US, whether a student, household size, individual and household income, education level, job status (whether part time, whether have another job), and subjective health status.

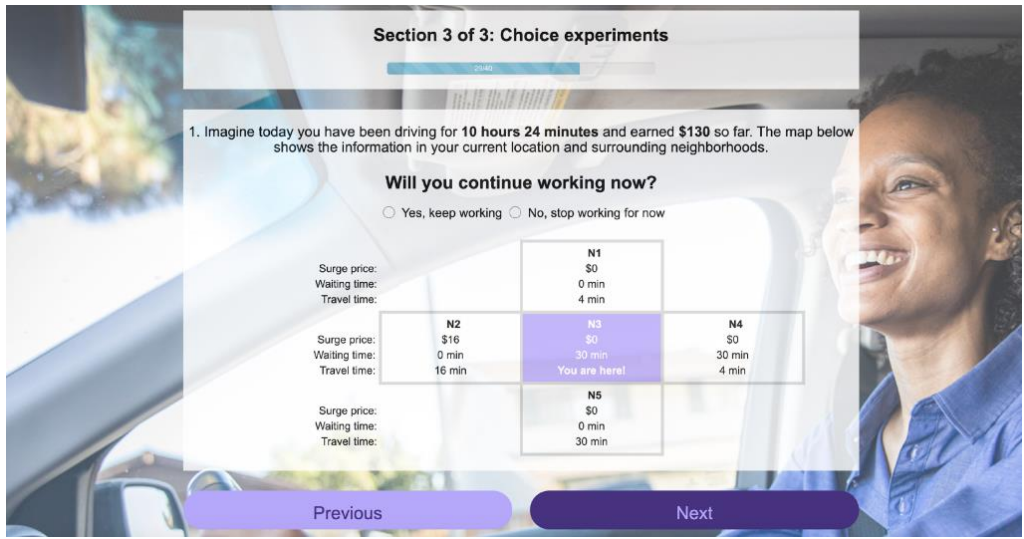

**Fig. 1 An example of the choice experiments: working time choice**

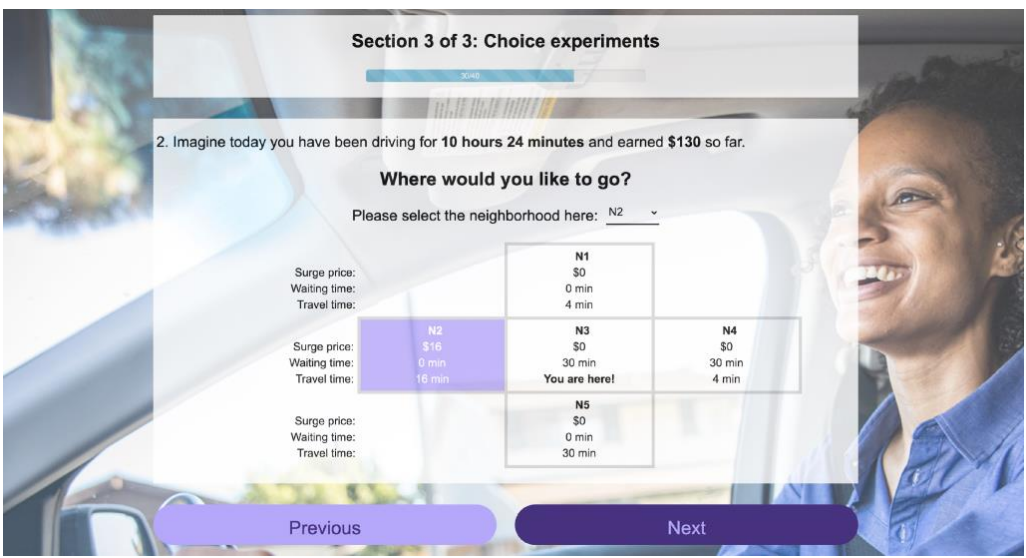

**Fig. 2 An example of the choice experiments: relocation choice**

**Table 1 Descriptions of experimental variables**

| Level | ID | Attributes | Descriptions |
|---|---|---|---|
| Working time choice | 1 | Earnings | Cumulative daily earnings at the time the choice is made. Base value is the respondent's self-reported daily earnings target. If the respondent did not report a daily earnings target, we assume the base value to be $100. |
| | 2 | Working time | Cumulative daily working hours at the time the choice is made. Base value is the respondent's self reported target working hours. If the respondent did not report a target number of working hours, we assume the base value to be 8 hours. |
| Relocation choice | 3 | Ridehailing demand (current neighborhood) | Represents the demand for ridehailing services in the current neighborhood. Measured jointly by *surge price ($)* and *average waiting time for a trip request if they decline the current one (minutes)*. Higher demand areas have lower waiting times and higher surge prices. |
| | 4-7 | Ridehailing demand (nearby neighborhoods) | Represents the demand for ridehailing services in each of the four nearby neighborhoods. Measured jointly by *surge price ($)* and *average waiting time for a trip request if they decline the current one (minutes)*. Higher demand areas have lower waiting times and higher surge price. |
| | 8-11 | Relocation time | The time to drive from the current neighborhood to each of the four nearby neighborhoods, in minutes. |

**Table 2 Experimental attribute levels**

| ID | Attributes | | Levels | | | | | | | | |
|---|---|---|---|---|---|---|---|---|---|---|---|
| 1 | Earnings Multiplier | | 0.4 | | 0.7 | | 1 | | 1.3 | | 1.6 |
| 2 | Working time multiplier | | 0.4 | | 0.7 | | 1 | | 1.3 | | 1.6 |
| 3-7 | Ridehailing demand | Surge price ($) | 16 | 12 | 8 | 4 | 0 | 0 | 0 | 0 | 0 |
| | | Average waiting time (min) | 0 | 0 | 0 | 0 | 0 | 8 | 16 | 24 | 30 |
| 8-11 | Relocation time (min) | | 4 | | 8 | | 12 | | 16 | | 20 |

### 3.2 Data quality and cleaning

To improve data quality, we adopted multiple screening and quality check questions as follows. First, before entering the choice scenario questions, survey respondents were asked two comprehension check questions. For each comprehension check, we asked the respondent to choose the correct meaning of a highlighted number in a choice scenario. Every respondent had two chances to answer each of the comprehension questions. They were only allowed to proceed if the answers were correct, otherwise the survey was terminated. Second, we included one question to check respondents' attention. The respondents were told "this is an attention check" and were asked to choose the neighborhood with given surge price, waiting time, and travel time in the middle of choice scenario questions. Finally, we applied a logic check in the survey. In the background information section, we asked respondents their annual income (both household and individual). Each respondent's annual household income should be equal or larger than their individual income. We removed responses failing either the attention or logic check.

For modeling purposes, we also removed respondents who had missing data or chose "prefer not to answer" in independent variables mentioned in Section 4. Respondents who completed 0 trips in the prior week were also removed.

After the data quality checks and cleaning, a total of 181 respondents and 1,067 choice responses were retained for analysis.

## 4. Model development

We adopted a mixed logit model to investigate ridehailing drivers' working time and relocation choices jointly. The model has two levels: working time choice and relocation choice. In the working time choice level, we have two alternatives: stop working and continue working. We choose whether to continue or stop working as the dependent variable following the approach of two previous studies (Farber 2008; Farber 2005). After the driver chooses to continue working, s/he can choose to stay in the same place or go to other places. If s/he decides to relocate, s/he can choose to go to one of the nearby 4 neighborhoods.

The strengths of a mixed logit model are twofold: for one, it can address the unobserved correlations between alternatives in the choice set with higher flexibility than nested logit. Second, it can also account for repeated measures. The modeling structure is shown in Figure 3. The mixed logit model is estimated in Biogeme by maximizing the simulated likelihood (Bierlaire 2020).

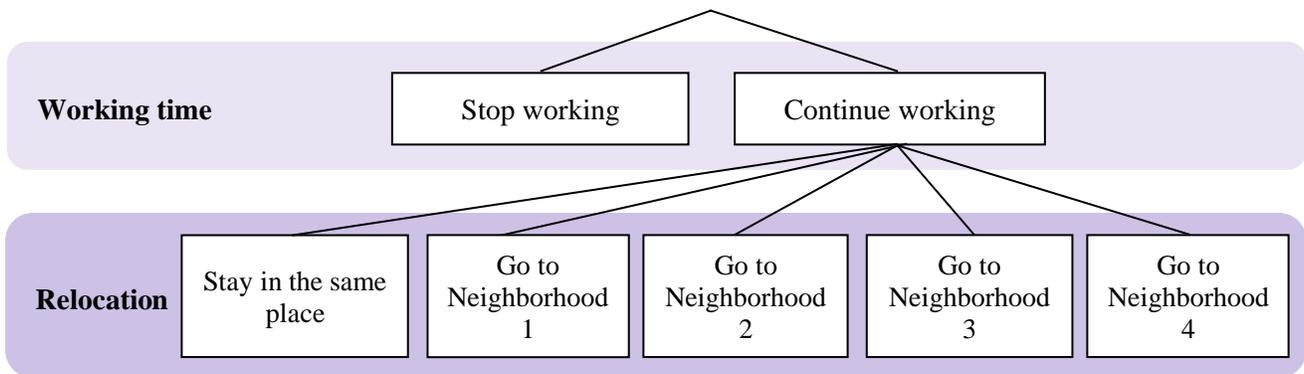

**Fig. 3 Choice structure for working time and relocation choice**

### 4.1 Working time choice

The utility of driver *i* choosing to stop working is written in Table 3. This model structure helps distinguish the effects of different types of drivers.

**Table 3 The utility function of stop working**

| Driver Type | Utility function | Driver Indicator | ASC | Earnings target | Time target | Working time | Earning rate | Covariates | Driver Random Effect | Error Term |
|---|---|---|---|---|---|---|---|---|---|---|
| No targets | $U_{i-stop} =$ | $D_{no}($ | $\alpha_{no} +$ | | | $\delta_{no}W_{ij} +$ | $\theta_{no}R_{ij}) +$ | $\eta_s X_i +$ | $\mu_i +$ | $\varepsilon_{ij}$ |
| Earnings Target | | $D_{et}($ | $\alpha_{et} +$ | $\beta_{et}E_{ij} +$ | | $\delta_{et}W_{ij} +$ | $\theta_{et}R_{ij}) +$ | $\eta_s X_i +$ | $\mu_i +$ | $\varepsilon_{ij}$ |
| Time Target | | $D_{tt}($ | $\alpha_{tt} +$ | | $\gamma_{tt}T_{ij} +$ | | $\theta_{tt}R_{ij}) +$ | $\eta_s X_i +$ | $\mu_i +$ | $\varepsilon_{ij}$ |
| Both Targets | | $D_{bt}($ | $\alpha_{bt} +$ | $\beta_{bt}E_{ij} +$ | $\gamma_{bt}T_{ij} +$ | | $\theta_{bt}R_{ij}) +$ | $\eta_s X_i +$ | $\mu_i +$ | $\varepsilon_{ij}$ |

Notations:

$D_{no}$, $D_{et}$, $D_{tt}$, and $D_{bt}$ are four dummy variables, representing *no-target drivers*, *earnings-target drivers*, *time-target drivers*, and *both-target drivers*. The descriptions of driver types are shown in Table 4.

$E_{ij}$ is a vector of earnings target variables indicating the category of driver $i$'s daily earnings in choice scenario $j$ relative to their daily earnings target. $\beta_{et}, \beta_{bt}$ are vectors of the coefficients of $E_{ij}$ for *earnings-target drivers* and *both-target drivers*.

$T_{ij}$ is a vector of working time target variables indicating the category of driver $i$'s daily working hours in choice scenario $j$ relative to their daily earnings target. $\gamma_{tt}, \gamma_{bt}$ are vectors of the coefficient of $T_{ij}$ for *time-target drivers* and *both-target drivers*.

$W_{ij}$ is the cumulative working time of driver $i$ in choice scenario $j$ at the time the choice is being made. $\delta_{no}, \delta_{et}$ are the coefficients of $W_{ij}$ for *no-target drivers* and *earnings-target drivers*. We only include $W_{ij}$ for *no-target drivers* and *earnings-target drivers* because they are highly correlated with time target variables for *time-target drivers* and *both-target drivers*.

$R_{ij}$ is the earning rate of driver $i$ in choice scenario $j$. $\theta_{no}, \theta_{et}, \theta_{tt}, \theta_{bt}$ are coefficients of $R_{ij}$ for *no-target drivers*, *earnings-target drivers*, *time-target drivers*, and *both-target drivers* respectively.

$X_i$ is a vector of socio-demographics including gender, age, number of children, income, and job status. $\eta_s$ is a vector of coefficients of $X_i$. The specific variables are shown in Tables 4.

$\mu_i$ is an individual-specific random component assumed to be distributed as $N\{0, \sigma^2\}$ where $\sigma$ is the variance component.

$\varepsilon_{ij}$ is a random error term assumed to be independently and identically Gumbel distributed.

The utility of driver $i$ choosing to continue working, is the expected utility of his/her available (re)location options, is written in *Eq. (1)*.

$$U_{i-continue} = \lambda ln \sum_{r \in m(i)} e^{(V_{ir}+b_i)*\frac{1}{\lambda}} + \vartheta_i \quad Eq.\ (1)$$

Where $m(i)$ is a set of available location choices (including staying in the same place and relocating to one of the nearby neighborhoods).

$\lambda$ is the scale parameter and it captures the unobserved similarities between alternatives. It ranges from 0 to 1; closer to 0 means higher unobserved correlations between the alternatives and closer to 1 means less correlations in the unobserved component of utility.

$V_{ir}$ is the systematic utility of driver $i$ choosing relocation choice $r$.

$\vartheta_{ir}$ is a random error term assumed to be independently and identically Gumbel distributed.

### 4.2 Relocation choice

The utility of driver $i$ choosing relocation choice $r$ is written as *Eq. (2)*.

$$U_{ir} = V_{ir} + b_i + \zeta_{ir} \quad Eq.\ (2)$$

$V_{ir}$, the systematic utility of driver $i$ choosing relocation choice $r$ is written as *Eq. (3)*.

$$V_{ir} = \rho S_{ir} + \omega A_{ir} + \tau L_{ir} + \phi_i \quad Eq.\ (3)$$

Where $r$ can be the current neighborhood the driver is at, or one of the nearby 4 neighborhoods.

$S_{ir}$ is surge price of neighborhood $r$. $\rho$ is the coefficient of $S_{ir}$.

$A_{ir}$ is the average waiting time of neighborhood $r$. $\omega$ is the coefficient of $A_{ir}$.

$L_{ir}$ is the relocation time from the current neighborhood to neighborhood $r$. $L_{ir}$ does not exist if driver $i$ chooses to stay in the current neighborhood. $\tau$ is the coefficient of $L_{ir}$.

$b_i$ is an individual-specific random component.

$\phi_i$ are the random effects used to account for unobserved correlations, including a vector of independent, standard normal distributed, random variables (continue working and relocating to other neighborhoods).

$\zeta_{ir}$ is a random error term assumed to be independently and identically Gumbel distributed.

To estimate the random effects, 500 Halton draws taken from a normal distribution were used.

**Table 4 Variable descriptions**

| Variable | Definition |
|---|---|
| **Working time choice** | |
| *Working characteristics* | |
| *Driver types* | |
| No-target drivers ($D_{no}$) | Dummy variable. If the respondent has neither working time target nor earnings target, 1, else, 0 |

| Earnings-target drivers ($D_{et}$) | Dummy variable. If the respondent has an earnings target but no working time target, 1, else, 0 |
|---|---|
| Time-target drivers ($D_{tt}$) | Dummy variable. If the respondent has a working time target but no earnings target, 1, else, 0 |
| Both-target drivers ($D_{bt}$) | Dummy variable. If the respondent has both working time target and earnings target, 1, else, 0 |
| *Earnings target ($E_{ij}$) (reference: Earnings = target)* | |
| Earnings = 0.4*target | Dummy variable. If accumulated earnings divided by the respondent's earnings target equals 0.4, 1, else, 0. Only applicable to earnings-target drivers and both-target drivers |
| Earnings = 0.7*target | Dummy variable. If accumulated earnings divided by the respondent's earnings target equals 0.7, 1, else, 0. Only applicable to earnings-target drivers and both-target drivers |
| Earnings = 1.3*target | Dummy variable. If accumulated earnings divided by the respondent's earnings target equals 1.3, 1, else, 0. Only applicable to earnings-target drivers and both-target drivers |
| Earnings = 1.6*target | Dummy variable. If accumulated earnings divided by the respondent's earnings target equals 1.6, 1, else, 0. Only applicable to earnings-target drivers and both-target drivers |
| *Working time target ($T_{ij}$) (reference: working time = target)* | |
| Working time = 0.4*target | Dummy variable. If accumulated working hours divided by the respondent's working time target equals 0.4, 1, else, 0. Only applicable to time-target drivers and both-target drivers |
| Working time = 0.7*target | Dummy variable. If accumulated working hours divided by the respondent's working time target equals 0.7, 1, else, 0. Only applicable to time-target drivers and both-target drivers |
| Working time = 1.3*target | Dummy variable. If accumulated working hours divided by the respondent's working time target equals 1.3, 1, else, 0. Only applicable to time-target drivers and both-target drivers |
| Working time = 1.6*target | Dummy variable. If accumulated working hours divided by the respondent's working time target equals 1.6, 1, else, 0. Only applicable to time-target drivers and both-target drivers |
| Working time ($W_{ij}$) | Accumulated working hours, in hours. Only applicable to no-target drivers and earnings-target drivers |
| Earning rate (earnings/working time) ($R_{ij}$) | Accumulated daily earnings/accumulated working hours |
| *Socio-demographics ($X_i$)* | |
| Age | The age of the respondent. 18 – 39: 0 40 – 64: 1 >=65: 2 |
| Female | If 1, gender = female, 0, otherwise |
| Num. of children | Number of children (under 18 years old) in the household |

| Above median income | If household income equal or higher than King County median household income in 2019 ($99,158) (U.S. Census Bureau 2021), 1; else, 0. |
|---|---|
| Another job | Whether the respondent has another job other than the ridehailing driver<br>0: has another job, full-time (35+ hours/week)<br>1: has another job, part-time (fewer than 35 hours/week)<br>2: doesn't have another job |
| **Relocation choice** | |
| Surge price ($S_{ir}$) | An additional surge amount to the trip fare if go to other places, in dollars |
| Average waiting time ($A_{ir}$) | Expected waiting time until next request if the respondent reject the current one if go to other places, in minutes |
| Relocation time ($L_{ir}$) | Driving time to the nearby neighborhood, in minutes |

## 5. Results
### 5.1 Descriptive analysis
#### 5.1.1 Socio-demographics and employment characteristics

The socio-demographics and employment characteristics are shown in Table 5. Compared to the general population, ridehailing drivers in our sample have some unique characteristics. They are overwhelmingly male (96%), Black American (75%), and born outside the US (95%). The finding is consistent with a report by City of Seattle, that 93% of TNC drivers identified themselves as non-white (City of Seattle 2020). Since the response rate to our survey was very high (~80%), we are confident that it is fairly representative of ridehailing drivers picking up passengers at SeaTac airport (all of whom pass through the waiting lot).

**Table 5** Data characteristics (N = 181)

| Characteristic | This sample | Characteristics | This sample |
|---|---|---|---|
| *What is your gender?* | | *Are you currently a student?* | |
| Male | 96% | Yes, full time (35+ hours/week) | 3% |
| Female | 4% | Yes, part time (<35 hours/week) | 7% |
| *What is your race?* | | No, not a student | 90% |
| White | 7% | *Are you of Hispanic, Latino, or Spanish origin?* | |
| Asian | 11% | Yes, Hispanic origin | 3% |
| African American | 75% | No, non-Hispanic origin | 94% |
| Another | 5% | Prefer not to answer | 3% |
| Prefer not to answer | 2% | *Were you born in the United States?* | |
| *What is your age?* | | Yes, born in the US | 4% |
| 18-39 | 51% | No, born outside the US | 95% |
| 40-64 | 45% | Prefer not to answer | 1% |
| 65 and above | 4% | *Do you have another job besides ride-sharing driver?* | |
| *What is the highest degree or level of school you have completed?* | | Yes, full time (35+ hours/week) | 9% |
| Less than high school | 5% | Yes, part time (<35 hours/week) | 12% |
| High school | 46% | No, no another job | 79% |
| College | 44% | *Which category best describes your household income before taxes from the last calendar year?* | |
| Graduate school or higher | 5% | Less than $10,000 | 3% |
| *How long have you been an active ride-sharing driver? "Active" means the time between two rides should be no longer than a month.* | | $10,000 to $14,999 | 2% |
| Less than 6 months | 5% | $15,000 to $19,999 | 3% |
| 6 months - less than 1 year | 5% | $20,000 to $24,999 | 9% |
| 1 year - less than 1.5 years | 2% | $25,000 to $34,999 | 10% |
| 1.5 years - less than 2 years | 7% | $35,000 to $49,999 | 20% |
| 2 years or more | 80% | $50,000 to $74,999 | 33% |
| Prefer not to answer | 1% | $75,000 to $99,999 | 11% |
| *How many people live in your household including yourself?* | | $100,000 to $199,999 | 7% |
| 1 | 29% | $200,000 to $249,999 | 1% |
| 2 | 17% | *Which ride-sharing companies are you driving for?* | |
| 3 | 13% | Uber only | 14% |
| 4 | 15% | Lyft only | 35% |
| 5 | 12% | Uber and Lyft | 51% |
| 6 | 6% | *Within your household, how many are children under the age of 18?* | |
| >=7 | 8% | 0 | 58% |
| *Do you start driving at around the same time every day?* | | 1 | 14% |
| Yes | 63% | 2 | 10% |
| No | 37% | 3 | 8% |
| *Do you stop driving at around the same time every day?* | | 4 | 4% |
| Yes | 50% | 5 | 3% |

| No | 50% | >=6 | 3% |

### 5.1.2 Driving characteristics

*Driver types*

Most respondents were *both-target drivers* (57%), with the remainder of the sample split fairly evenly among *no-target drivers* (14%), *time-target drivers* (17%), and *earnings-target drivers* (13%).

*Target characteristics*

Figure 4 shows the distributions of daily target earnings (Figure 4a) and working time (Figure 4b). The variability of driver earnings targets is large, ranging from $100 to $400. The mean earnings target is $221. The variability of daily working hour targets is also large across drivers, with a minimum of 4 hours and a maximum of 12 hours. The mean target is 9 hours. We compared the daily working hour target we collected with actual daily working time of taxi drivers (Agarwal et al. 2013; Farber 2008). They are very similar: New York taxi drivers' working hours per shift ranges from 7 to 9 hours (Farber 2008), and Singapore drivers mainly drive between 8 to 10 hours (Agarwal et al. 2013).

For *both-target drivers*, the working time targets and earnings targets are moderately correlated: the correlation coefficient is 0.39. This is consistent with the intuition that drivers who expect to work longer would also expect to earn more.

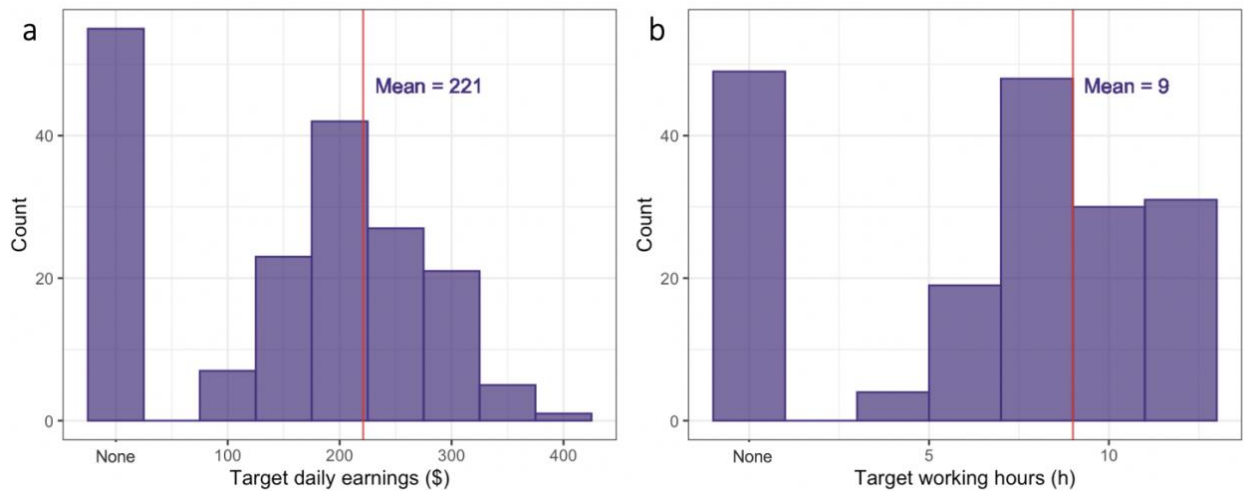

**Fig. 4 Target daily earnings and working hours**

### 5.2 Inferential analysis

Table 6 presents the mixed logit model estimation results for both working time and relocation choices. Key features of these results are examined in the sections that follow.

### 5.2.1 Working time choice

*Earning rate*

Overall, ridehailing drivers are more likely to continue working when their earning rate is higher. However, the magnitude and significance of this effect vary across different types of drivers. Earning rate has the largest impact for *no-target drivers*, followed by *time-target drivers* and *earnings-target drivers*. However, the effect of earning rate on the working time choices of *both-target drivers* is small and non-significant.

*Working time*

We only include the working time variable for *no-target drivers* and *earnings-target drivers*. For *time-target drivers* and *both-target drivers*, the working time variable is not included because its physical meaning is very similar to working time target variables. Overall, having already worked more hours in a day leads to a higher probability of stopping working. However, the effect is statistically significant only for *no-target drivers*.

*Working time target*

We only include working time target variables for *time-target drivers* and *both-target drivers*. The coefficients of the working time target variables are shown in Figure 5. The level "working time = target" is used as the reference level, and so its coefficient is fixed to 0 by definition. Both *time-target drivers* and *both-target drivers* are most likely to continue working when they are below their working time targets, and they become more likely to stop working after reaching the target.

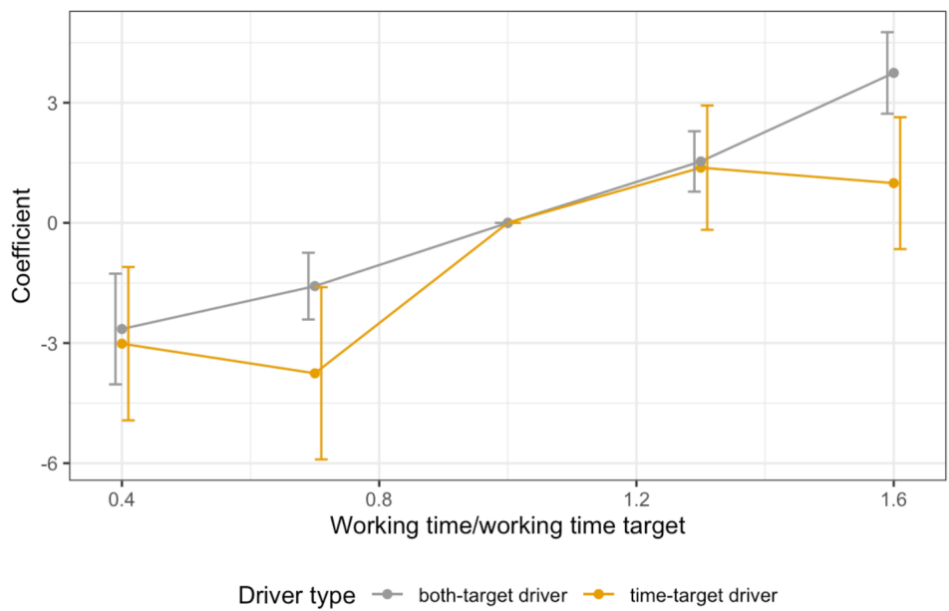

**Fig. 5 Working time target effect for *time-target drivers* and *both-target drivers***

*Note: error bar represents 95% confidence interval*

*Earnings target*

We only include earnings target variables for *earnings-target drivers* and *both-target drivers*. Both *earnings-target drivers* and *both-target drivers* are more likely to stop working when they reach 160% of their earnings target. However, the earnings target variables are all non-significant and thus should be interpreted with caution.

*Socio-demographics*

Female drivers and drivers with more children are more likely to continue working, all else equal. We found no significant differences in working time choice across age groups or other employment status.

### 5.2.2 Relocation choice

The ASC $_{stay\ in\ the\ same\ place}$ is positively associated with staying in the same place. This means that drivers are more likely to stay in the same neighborhood, everything else being equal. The result is consistent with a study by Henao and Marshall (Henao and Marshall 2019). They found that drivers earn more if they wait for the next trip request in the same place rather than driving to other places, unless they can save at least 30% of trip waiting time.

*Neighborhood characteristics*

Surge price is significant for both staying in the same place and relocating to other places, meaning that drivers prefer a neighborhood with a higher surge price.

Average trip waiting time has a negative impact on both staying in the same place and relocating to other places, meaning that drivers are less likely to work in a place where there is a long expected wait to get a trip request.

Relocation time is negatively associated with relocating to other places, suggesting that drivers are less likely to relocate to places when it takes a long time to drive there.

*Error components*

Both error components (continue working and relocate to other places) are significant, meaning that our assumption on the model structure is valid by capturing both unobserved correlations between different options involving continuing to work, and between different neighborhoods to which the driver could relocate.

**Table 6 Modeling results (N = 1067). Standard errors are in parentheses.**

| Variable | Drivers without targets (N = 140) | Drivers with time target only (N = 174) | Drivers with earnings target only (N = 144) | Drivers with both targets (N = 609) |
|---|---|---|---|---|
| **Working time choice: stop working** | | | | |
| *Working characteristics* | | | | |
| ASC *stop working* | 1.888 (1.637) | 3.853 (0.788)*** | 2.123 (1.836) | 2.629 (0.618)*** |
| Earning rate (earnings/working time) | -0.194 (0.059)*** | -0.090 (0.044)** | -0.041 (0.020)** | $-1.4 \times 10^{-5}$ (0.015) |
| Working time | 0.452 (0.129)*** | – | 0.188 (0.130) | – |
| *Earnings target (reference: Earnings = target)* | | | | |
| Earnings = 0.4*target | – | – | -0.152 (0.847) | -0.201 (0.509) |
| Earnings = 0.7*target | – | – | -0.572 (0.970) | -0.671 (0.455) |
| Earnings = 1.3*target | – | – | -0.667 (0.815) | -0.151 (0.426) |
| Earnings = 1.6*target | – | – | 1.741 (1.072) | 0.174 (0.506) |
| *Working time target (reference: working time = target)* | | | | |
| Working time = 0.4*target | – | -3.016 (0.977)*** | – | -2.650 (0.705)*** |
| Working time = 0.7*target | – | -3.756 (1.097)*** | – | -1.579 (0.425)*** |
| Working time = 1.3*target | – | 1.379 (0.792)* | – | 1.534 (0.385)*** |
| Working time = 1.6*target | – | 0.992 (0.840) | – | 3.744 (0.519)*** |
| *Socio-demographics* | | | | |
| Age (reference: 40~64) | | | | |
|   Young adults (18~39) | colspan across: -0.193 (0.376) | | | |
|   Older adults (>=65) | colspan across: 0.182 (0.898) | | | |
| Gender (reference: male) | | | | |
|   Female | colspan across: -3.383 (1.051)*** | | | |
| Num. of children | colspan across: -0.305 (0.109)*** | | | |

| | |
|---|---|
| Household annual income | |
|    Higher than city median | -0.909 (0.644) |
| Do you have another job? (reference: No, I don't have another job) | |
|    Yes, I have another job | -0.760 (0.560) |
| **Relocation choice: stay in the same place** | |
| *Neighborhood characteristics* | |
| ASC $_{stay\ in\ the\ same\ place}$ | 0.306 (0.157)* |
| Surge price | 0.184 (0.012)*** |
| Average trip waiting time | -0.038 (0.008)*** |
| Error component $_{continue\ working}$ | 1.935 (0.214)*** |
| **Relocation choice: relocate to other places** | |
| *Neighborhood characteristics* | |
| Surge price | 0.184 (0.012)*** |
| Average trip waiting time | -0.038 (0.008)*** |
| Relocation time $_{relocate\ to\ other\ places}$ | -0.124 (0.012)*** |
| Error component $_{continue\ working}$ | 1.935 (0.214)*** |
| Error component $_{relocate\ to\ other\ places}$ | 0.466 (0.228)** |
| **Model performance** | |
| Log Likelihood | -970 |

*Note: \* significant at 0.1 level, \*\* significant at 0.05 level, \*\*\* significant at 0.01 level*

# 6. Surge price effect

In our model structure, surge price has a direct impact on relocation choices, and an indirect impact on working time choices. First, the characteristics of each neighborhood determine the utility of relocating to that neighborhood. In turn, the expected utility over all neighborhoods affects the utility of continuing to work. For example, increasing surge price of the current neighborhood increases utility of staying in the same neighborhood. The higher expected utility at the relocation level in turn increases the utility of continuing to work.

Via a sensitivity analysis, we can understand how surge price would impact both working time and relocation choices. We show the surge price elasticity of drivers stopping working (Figure 6) and relocating (Figure 7). We vary the surge price in the current neighborhood and a nearby neighborhood[3] and observe the percentage change of the market share of drivers that choose to stop working (Figure 6) and to stay in the same place/go elsewhere (Figure 7).

## 6.1 Working time choice

Overall, increasing the surge price in either the current or a nearby neighborhood encourages drivers to continue working (Figure 6). This pattern is consistent among all types of drivers. *Earnings-target drivers* are most sensitive to the change of surge price, followed by *both-target drivers*, *time-target drivers*, and *no-target drivers*. This suggests that increasing the same amount of surge price would encourage a larger share of *earnings-target drivers* to continue working compared with other types of drivers.

However, changing the surge price in the current neighborhood almost doubles the impact decisions to stop working, compared with changing the surge in a nearby neighborhood. The result suggests that drivers would be more likely to stay in the same place than going elsewhere if increasing the same amount of surge price in the current neighborhood and a nearby one.

---

[3] Since the nearby neighborhoods are unlabeled alternatives, the effect is the same changing any of nearby neighborhoods. We choose neighborhood 1 for convenience.

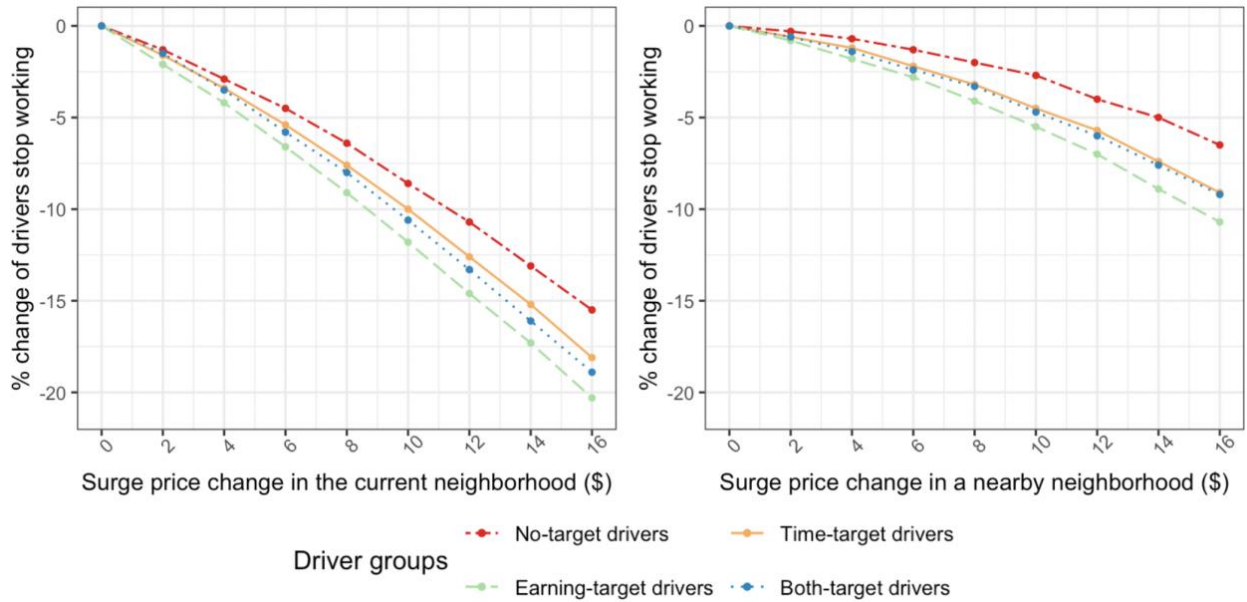

**Fig. 6 Predicted working time choice under future scenarios**

### 6.2 Relocation choice

Increasing the surge price in the current neighborhood encourages drivers to stay where they are. The horizontal axis in Figure 7 represents the surge price change, while the vertical axis represents the absolute difference between current percentage and original percentage of drivers relocating. For example, increasing surge price in the current neighborhood by $16 would first increase the share of drivers continuing to work by 19%. Among those who continue working, 40% more drivers would choose to stay in the same place than before, while 5% less drivers would choose to relocate to another neighborhood on average. This result suggests that increasing the surge price in the current neighborhood increases driver supply in that neighborhood in two ways: first, it encourages drivers to continue working; second, it discourages drivers from relocating to other neighborhoods.

If we increase surge price in a nearby neighborhood, we observe more drivers going there. For example, increasing surge price in a nearby neighborhood by $16 would first encourage 10% of drivers to continue working. Among those who continue working, the surge price would encourage 25% of drivers to go there; 7% drivers who otherwise would have stayed in their current neighborhood, 2% drivers who would have relocated to a different neighborhood would instead go to the neighborhood with the $16 surge price.

Changing the surge price in the current neighborhood has larger impact on drivers' relocation choice than changing it in a nearby neighborhood.

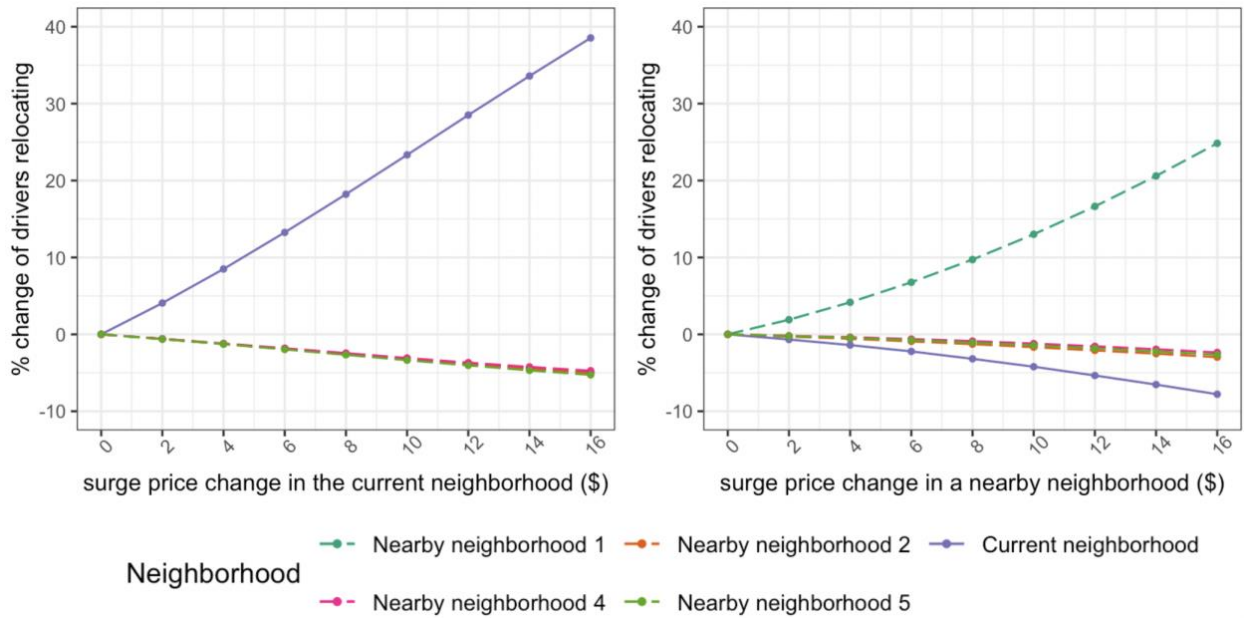

**Fig. 7 Predicted relocation choice under future scenarios**

## 7. Conclusions

Using 1,067 stated choice responses, our study first identified four types of ridehailing drivers based on their working time and earnings targets, then jointly modeled their working time choice and relocation choices. Based on the modeling results, we simulated the impact of surge price on both working time and relocation choices.

*Working time choice*

Our finding mainly echoes the neo-classic theory of driver labor supply: all types of drivers choose to continue working as their earning rate increases (Lucas and Rapping 1969). We see little evidence of an earnings threshold effect that the reference dependent theory postulates (Tversky and Kahneman 1991), even when the driver reported having an earnings target in the survey. However, there is evidence of reference dependent behavior with respect to the working time variable: *time-target drivers* and *both-target drivers* are less likely to continue working after hitting their working time targets.

*Relocation choice*

Relocation choice is mainly impacted by surge price, average trip waiting time, and relocation time: higher surge price in the current neighborhood encourages the driver to stay, while higher surge price in a nearby neighborhood attracts the driver to relocate. Drivers are more likely to stay in a neighborhood where the average trip waiting time is low. Longer relocation time discourages drivers to relocate. In addition, drivers are more likely to stay in the same place, everything else being equal.

*Future research*

There are several future directions in which our work can be improved: First, future studies might consider including drivers from other channels (e.g., hailing rides) to provide a more representative view of ridehailing drivers. The sample in our study only comes from the airport waiting area. Future studies might consider including drivers from other channels (e.g., hailing rides) to provide a more representative view of ridehailing drivers. However, our experience shows that ridehailing drivers are a notoriously difficult community to reach and recruit for survey research. Second, future efforts should identify and consider other important factors such as pay rates of different cities to have a better understanding of the relationship between trip features and drivers' behavior.

## CONTRIBUTION STATEMENT

**Yuanjie (Tukey) Tu:** Conceptualization, Survey design, Data collection, Methodology, Formal analysis, Software, Writing - Original draft preparation; **Moein Khaloei:** Survey design, Data collection; **Natalia Zuniga-Garcia:** Writing – Review & Editing, Supervision; **Don MacKenzie:** Conceptualization, Survey design, Methodology, Writing – Review & Editing, Supervision, Project administration, Funding acquisition. All authors read, commented, and approved the final version of the manuscript.

## ACKNOWLEDGEMENT

The work described here was sponsored by the U.S. Department of Energy (DOE) Vehicle Technologies Office (VTO) under the Systems and Modeling for Accelerated Research in Transportation (SMART) Mobility Laboratory Consortium, an initiative of the Energy Efficient Mobility Systems (EEMS) Program. The submitted manuscript has been created by the University of Washington and the UChicago Argonne, LLC, Operator of Argonne National Laboratory (Argonne). Argonne, a U.S. Department of Energy Office of Science laboratory, is operated under Contract No. DE-AC02-06CH11357. The U.S. Government retains for itself, and others acting on its behalf, a paid-up nonexclusive, irrevocable worldwide license in said article to reproduce, prepare derivative works, distribute copies to the public, and perform publicly and display publicly, by or on behalf of the Government.

## DECLARATION OF INTERESTS

The authors declare that they have no known competing financial interests or personal relationships that could have appeared to influence the work reported in this paper.